\documentclass[aps,prl,preprint,superscriptaddress,showpacs,byrevtex]{revtex4}
\usepackage{graphicx} 
\usepackage{dcolumn}  

\graphicspath{{ps}}

\newcommand{\ee}{e^{+} e^{-}}
\newcommand{\bbar}{B\bar{B}}

\newcommand{\jp}{J/\psi}
\newcommand{\psip}{\psi '}

\newcommand{\pipi}{\pi^{+}\pi^{-}}
\newcommand{\piz}{\pi^{0}}
\newcommand{\ks}{K_{S}^0}

\newcommand{\ecp}{\eta_{c}(2S)}
\newcommand{\kpi}{K^-\pi^{+}}
\newcommand{\Mbc}{M_{\rm bc}}
\newcommand{\DE}{\Delta E}

\newcommand{\rt}{\rightarrow}
\newcommand{\etal}{\em et al.}

\begin{document}


\preprint{\vbox{ \hbox{   }
}}

\title{ \quad\\[0.5cm] Observation of a near-threshold $\omega\jp$ 
mass enhancement in exclusive 
$B \rt K \omega\jp$ decays }

\affiliation{Budker Institute of Nuclear Physics, Novosibirsk}
\affiliation{Chiba University, Chiba}
\affiliation{Chonnam National University, Kwangju}
\affiliation{University of Cincinnati, Cincinnati, Ohio 45221}
\affiliation{University of Frankfurt, Frankfurt}
\affiliation{Gyeongsang National University, Chinju}
\affiliation{University of Hawaii, Honolulu, Hawaii 96822}
\affiliation{High Energy Accelerator Research Organization (KEK), Tsukuba}
\affiliation{Institute of High Energy Physics, Chinese Academy of Sciences, Beijing}
\affiliation{Institute of High Energy Physics, Vienna}
\affiliation{Institute for Theoretical and Experimental Physics, Moscow}
\affiliation{J. Stefan Institute, Ljubljana}
\affiliation{Kanagawa University, Yokohama}
\affiliation{Korea University, Seoul}
\affiliation{Kyungpook National University, Taegu}
\affiliation{Swiss Federal Institute of Technology of Lausanne, EPFL, Lausanne}
\affiliation{University of Ljubljana, Ljubljana}
\affiliation{University of Maribor, Maribor}
\affiliation{University of Melbourne, Victoria}
\affiliation{Nagoya University, Nagoya}
\affiliation{Nara Women's University, Nara}
\affiliation{National Central University, Chung-li}
\affiliation{National United University, Miao Li}
\affiliation{Department of Physics, National Taiwan University, Taipei}
\affiliation{H. Niewodniczanski Institute of Nuclear Physics, Krakow}
\affiliation{Nihon Dental College, Niigata}
\affiliation{Niigata University, Niigata}
\affiliation{Osaka City University, Osaka}
\affiliation{Osaka University, Osaka}
\affiliation{Panjab University, Chandigarh}
\affiliation{Peking University, Beijing}
\affiliation{Saga University, Saga}
\affiliation{University of Science and Technology of China, Hefei}
\affiliation{Seoul National University, Seoul}
\affiliation{Sungkyunkwan University, Suwon}
\affiliation{University of Sydney, Sydney NSW}
\affiliation{Tata Institute of Fundamental Research, Bombay}
\affiliation{Toho University, Funabashi}
\affiliation{Tohoku Gakuin University, Tagajo}
\affiliation{Tohoku University, Sendai}
\affiliation{Department of Physics, University of Tokyo, Tokyo}
\affiliation{Tokyo Institute of Technology, Tokyo}
\affiliation{Tokyo Metropolitan University, Tokyo}
\affiliation{Tokyo University of Agriculture and Technology, Tokyo}
\affiliation{University of Tsukuba, Tsukuba}
\affiliation{Virginia Polytechnic Institute and State University, Blacksburg, Virginia 24061}
\affiliation{Yonsei University, Seoul}
  \author{S.-K.~Choi}\affiliation{Gyeongsang National University, Chinju} 
  \author{S.~L.~Olsen}\affiliation{University of Hawaii, Honolulu, Hawaii 96822} 
  \author{K.~Abe}\affiliation{High Energy Accelerator Research Organization (KEK), Tsukuba} 
  \author{K.~Abe}\affiliation{Tohoku Gakuin University, Tagajo} 
  \author{I.~Adachi}\affiliation{High Energy Accelerator Research Organization (KEK), Tsukuba} 
  \author{H.~Aihara}\affiliation{Department of Physics, University of Tokyo, Tokyo} 
  \author{Y.~Asano}\affiliation{University of Tsukuba, Tsukuba} 
  \author{S.~Bahinipati}\affiliation{University of Cincinnati, Cincinnati, Ohio 45221} 
  \author{A.~M.~Bakich}\affiliation{University of Sydney, Sydney NSW} 
  \author{Y.~Ban}\affiliation{Peking University, Beijing} 
% ...
\collaboration{The Belle Collaboration}
\noaffiliation

\begin{abstract}

We report the observation of a near-threshold enhancement in the 
$\omega\jp$ invariant mass distribution for exclusive 
$B\rt K \omega\jp$ decays.  
The results are obtained from a 253~fb$^{-1}$ data sample
that contains 275 million $B\bar{B}$ pairs that were
collected near the $\Upsilon(4S)$ resonance with the Belle detector at 
the KEKB asymmetric energy $\ee$ collider.
The statistical significance of the
$\omega\jp$ mass enhancement is estimated to be
greater than $8\sigma$.

\end{abstract}


\maketitle

\tighten

{\renewcommand{\thefootnote}{\fnsymbol{footnote}}}
\setcounter{footnote}{0}


Recently there has been a revival of interest in the possible
existence of mesons with a more complex structure
than the simple $q\bar{q}$ bound states of the original quark model.
There are long-standing predictions of four-quark $q\bar{q}q\bar{q}$ 
meson-meson resonance states~\cite{molecule} and for $q\bar{q}$-$gluon$ 
hybrid states~\cite{mandula}.  Searches for these types of particles
in systems that include a charmed-anticharmed quark pair ($c\bar{c}$) 
are particularly effective  because for at least some of these cases,
the states are expected to have clean experimental signatures 
as well as relatively narrow widths, thereby reducing the 
possibility of overlap with standard  
$c\bar{c}$ mesons.

$B$ meson decays are a prolific source of $c\bar{c}$
pairs and the large $B$ meson samples produced at $B$-factories are providing
opportunities to search for missing $c\bar{c}$ charmonium mesons as well
as more complex states.
From studies of $\ks\kpi$ systems produced in exclusive 
$B\rt K\ks\kpi$  decays with a 45~million $\bbar$ event sample, 
the Belle group  made the first  observation of the 
$\ecp$~\cite{skchoi_etacp}.  This state was subsequently
seen in two-photon reactions by other experiments~\cite{cleo_etacp}
and in the exclusive $\ee\rt\jp\ecp$ production process by 
Belle~\cite{pakhlov_etacp}.
With a larger sample of 152~million $\bbar$ events, Belle discovered
the $X(3872)$ as a narrow peak in the $\pipi\jp$ mass spectrum 
from exclusive $B\rt K\pipi\jp$ decays~\cite{skchoi_x3872}.
This observation has been confirmed by other 
experiments~\cite{CDF}.   The properties of the $X(3872)$ 
do not match well to any $c\bar{c}$ charmonium state~\cite{slo_krakow}.
This, together with the close proximity of the $X(3872)$ mass
with the $m_{D^0} + m_{D^{*0}}$ mass threshold, have led
a number of authors to interpret the $X(3872)$ as 
a $D^0\bar{D}^{*0}$ resonant state~\cite{tornqvist}.

In this Letter we report on a study of the $\omega\jp$ system produced 
in exclusive $B\rt K\omega\jp$ decays. We observe a large enhancement in the 
$\omega\jp$ mass distribution  near the $\omega\jp$ mass threshold.  These 
results are based on a 253~fb$^{-1}$ data sample that contains 275~million 
$\bbar$ pairs collected with the Belle detector.

The Belle experiment observes $B$ mesons produced by the KEKB
asymmetric energy $e^+e^-$ collider~\cite{KEKB}.
KEKB operates at the $\Upsilon(4S)$ resonance
($\sqrt{s}=10.58$~GeV) with
a peak luminosity of $1.39\times 10^{34}~{\rm cm}^{-2}{\rm s}^{-1}$. 

The Belle detector is a large-solid-angle magnetic
spectrometer that consists of a silicon vertex detector,
a 50-layer cylindrical drift chamber (CDC), an array of
aerogel threshold Cherenkov counters (ACC),  
a barrel-like arrangement of time-of-flight 
scintillation counters (TOF), and an electromagnetic calorimeter
(ECL) comprised of CsI(Tl) crystals  located inside
a superconducting solenoid coil that provides a 1.5~T
magnetic field.  An iron flux-return located outside of 
the coil is instrumented to detect $K_L$ mesons and to identify
muons (KLM).  
The Belle detector is described in ref.~\cite{Belle}.


We select events of the type $B\rt K\pipi\piz\jp$, where we use
both charged and neutral kaons~\cite{conj}.  We use the  same charged kaon,
pion and $\jp$ requirements as used in ref.~\cite{skchoi_x3872}.
For neutral kaons we use $\pipi$ pairs with invariant mass
within 15~MeV of $m_{K_S}$ and a displaced vertex that is
consistent with $\ks\rt\pipi$ decay. 
We identify a $\piz$ as a $\gamma\gamma$ pair that fits the
$\pi^0\rt\gamma\gamma$ hypothesis with $\chi^2<6$.  We further require the 
energy asymmetry 
$|E_{\gamma_1}-E_{\gamma_2}|/|E_{\gamma_1}+E_{\gamma_2}|<0.9$ and
the $\pi^0$ laboratory-frame momentum to be greater
than 180~MeV.  Events with a $\pipi\jp$
invariant mass within $3\sigma$ of $m_{\psip}$ are rejected in
order to eliminate $B\rt K\pi^0\psip$; $\psip\rt\pipi\jp$
decays. The level of $\ee\rt q\bar{q}$ ($q=u,d,s~{\rm or}~c$-quark)
continuum events in the sample is reduced by the 
requirements  $R_2 < 0.4$, where $R_2$ is the normalized
Fox-Wolfram moment~\cite{fox}, and $|\cos\theta_B|<0.8$, where
$\theta_B$ is the polar angle of the $B$-meson direction
in the center-of-mass system (cms).

At the $\Upsilon(4S)$, $B\bar{B}$ meson pairs are
produced with no accompanying particles. As a result, each
$B$ meson has a total cms 
energy that is equal to $E_{\rm beam}$, the cms beam energy.  
We identify $B$ mesons using the beam-constrained mass 
$\Mbc=\sqrt{E_{\rm beam}^2 - p_B^2}$ and the energy difference
$\Delta E = E_{\rm beam} - E_B$, where $p_B$ is the vector sum of the 
cms momenta of the $B$ meson decay products and $E_B$ is their
cms energy sum.  
For the final state used in this analysis,
the experimental resolutions for $\Mbc$ and $\DE$ are
approximately 3~MeV and 13~MeV, respectively.

We select events with 
$\Mbc>5.20$~GeV and $|\DE|<0.2$~GeV for
further analysis.  For events 
with multiple $\pi^0$ entries in this region, 
we select the $\gamma\gamma$ combination
with the best $\chi^2$ value for the $\pi^0\rt\gamma\gamma$ 
hypothesis.
Multiple entries caused by multiple charged particle
assignments are small ($\sim4\%$) and are tolerated.
The signal region is defined as 5.2725~GeV$< \Mbc <$5.2875 GeV
and $|\DE |< $ 0.030 GeV, which is about $\pm 2.5\sigma$ 
from the central values.  We identify three-pion
combinations with $0.760~{\rm GeV}<M(\pipi\pi^0)<0.805$~GeV as 
candidate $\omega$ mesons. 
To suppress events of the type $B\rt K_X\jp$; $K_X \rt K\omega$,  
where $K_X$  denotes strange meson resonances such as  
$K_1(1270)$, $K_1(1400)$, and $K^*_2(1430)$ that are 
known to decay to $K\omega$, we restrict our
analysis to events in the region $M(K\omega)>1.6$~GeV.

\begin{figure}[htb]
\includegraphics[width=0.75\textwidth]{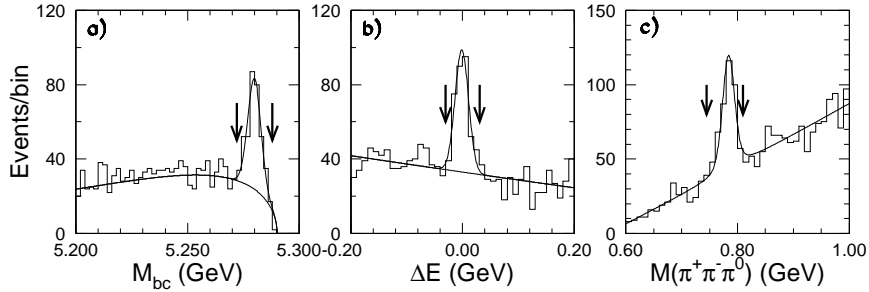}
\caption{ {\bf (a)} $\Mbc$ distributions for $B\rt K\omega\jp$
candidates in the $\DE$ and $\omega\rt\pipi\pi^0$
 signal regions. {\bf (b)} $\DE$ distribution for
events in the $\Mbc$ and $\omega$ signal regions. 
{\bf (c)} $M(\pipi\pi^0)$ distribution
for events in the $\Mbc$ and $\DE$ signal regions. 
The  curves are the results of fits described in the text;
the arrows indicate the signal region for each
quantity.}
\label{fig:mb_de_3pi}
\end{figure}

Figure~\ref{fig:mb_de_3pi}(a) shows the $\Mbc$ distribution
for selected events that are in the $\DE$ and $\omega$
signal regions.  
The curve in the figure is the result
of a binned likelihood fit that uses a single Gaussian
for the signal and an ARGUS function~\cite{ARGUS} for the background.  
The fit gives a signal yield of $219\pm 23$ events.
Figure~\ref{fig:mb_de_3pi}(b) shows the $\DE$ distribution
for events in the $\Mbc$ and $\omega$ signal regions.
Here the curve is the result of a fit that
represents the signal with a single Gaussian and
the background with a first-order polynomial.
The signal yield is $196 \pm 21$ events.
Figure~\ref{fig:mb_de_3pi}(c) shows the 
$M(\pipi\pi^0)$ distribution for all events
in the $\Mbc$-$\DE$ signal region.  The peak is
well fitted with a Breit-Wigner with the mass
and width of the $\omega(780)$, broadened by an
experimental resolution of 8~MeV. Here
the signal yield is $204 \pm 20$ events.
The reasonable consistency in the signal
yield from all three distributions
indicates that $K\omega\jp$ is the dominant
component of $B\rt K\pipi\piz\jp$ decays with
$M(3\pi)$ in the $\omega$ mass range and
$M(K\omega)>1.6$~GeV. 
The arrows in the figures indicate the signal regions 
for the plotted quantity.

\begin{figure}[htb]
\includegraphics[width=0.4\textwidth]{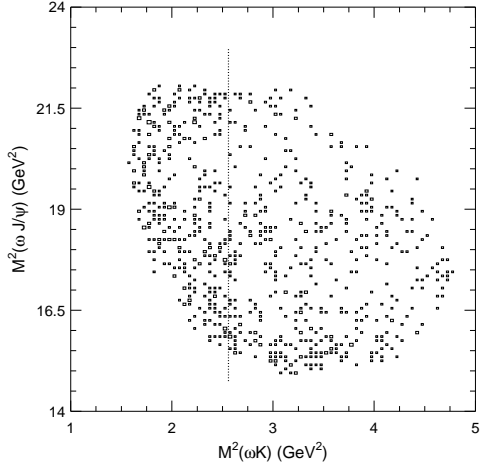}
\caption{ Dalitz-plot distribution for 
$B\rt K\omega\jp$ candidate events. The dotted line
indicates the boundary of the $M(K\omega)>1.6$~GeV
selection requirement.}
\label{fig:dalitz}
\end{figure}

Figure~\ref{fig:dalitz} shows the Dalitz plot of
$M^2(\omega\jp)$ (vertical) $vs$ $M^2(\omega K)$
(horizontal) for $B\rt K\omega\jp$ candidates
in the signal regions.  
Here the $M(K\omega)>1.6$~GeV requirement has been
relaxed. The clustering of events near the left side 
of the plot are attributed to $B\rt K_X\jp$; $K_X \rt K\omega$
decays.   There is an additional clustering of events with
low $\omega\jp$ invariant masses  near the bottom
of the Dalitz plot.   This clustering is the
subject of the analysis reported here.


\begin{figure}[htb]
\includegraphics[width=0.6\textwidth]{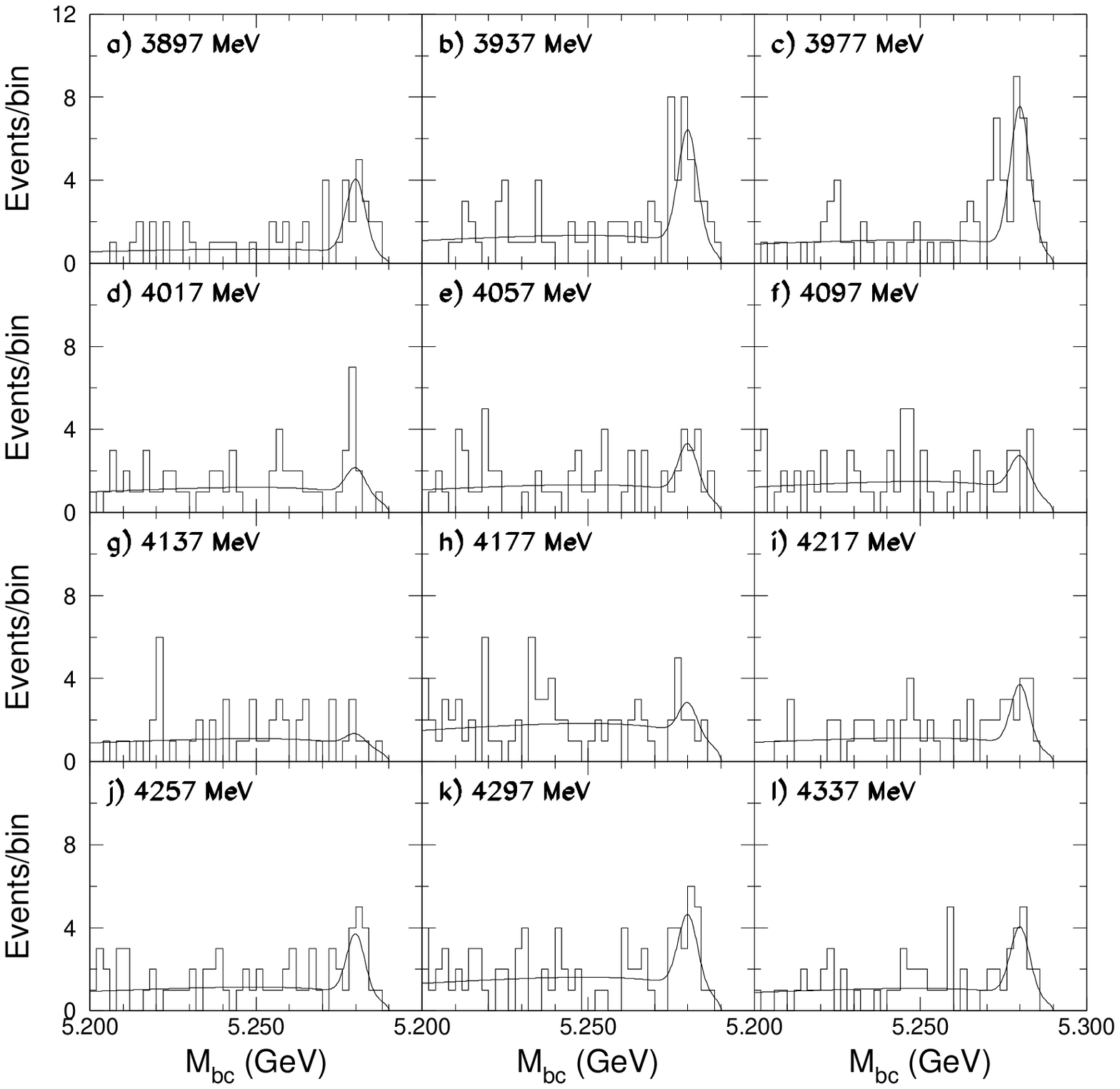} 
\caption{\label{fig:mb_12box}
$\Mbc$ distributions for $B^-\rt K^-\omega\jp$
candidates in the $\DE$
signal region for 40~MeV-wide $\omega\jp$
invariant mass intervals.  
The curves are the results of
fits described in the text.
}
\end{figure}

The fits to the $\Mbc$ and $\DE$ distributions
of Figs.~\ref{fig:mb_de_3pi}(a) and (b)
indicate that about half of the entries with
$M(\omega K)>1.6$~GeV in the Dalitz plot 
of Fig.~\ref{fig:dalitz}
are background.  To determine the level of
$B\rt K\omega\jp$ signal events, we bin
the data into $40$~MeV-wide bins of $M(\omega\jp)$
and fit for $B$ meson signals.
The histograms in Figs.~\ref{fig:mb_12box}(a)-(l)
show the $\Mbc$ distributions for the twelve lowest
$M(\omega\jp)$ bins for events in the $\DE$
and $\omega$ signal regions.  Here
there are distinct $B\rt K\omega\jp$ signals for 
low $\omega\jp$ invariant mass bins, especially 
those covered by Figs.~\ref{fig:mb_12box}(b) and (c).
We establish the $B\rt K\omega\jp$ signal level
for each $M(\omega\jp)$ bin by performing
binned one-dimensional fits to the $\Mbc$ and $\DE$ 
distributions for events in that interval  using the
same signal and background functions 
that are used to fit the integrated 
distributions of  Figs.~\ref{fig:mb_de_3pi}(a) and (b). 
For these  fits, the peak positions, 
resolution values and background shape parameters are 
all fixed at the values that are determined 
from the fits to the integrated 
distributions, and the areas of the
$\Mbc$ and $\DE$ signal functions are
constrained to be equal.
The curves in Fig.~\ref{fig:mb_12box} 
indicate the results of the $\Mbc$ fits.


\begin{figure}[htb]
\includegraphics[width=0.6\textwidth]{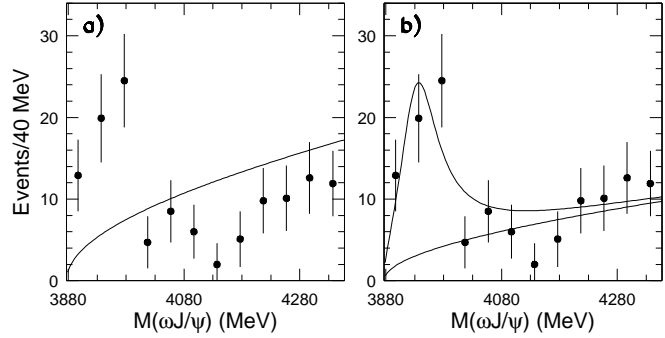} 
\caption{ $B\rt K\omega\jp$ signal yields
$vs$ $M(\omega\jp).$  The curve in {\bf (a)}
indicates the result of a fit that includes
only a phase-space-like threshold function.
The curve in {\bf  (b)}
shows the result of a fit that includes an
$S$-wave Breit-Wigner resonance term.}
\label{fig:slice_fits}
\end{figure}

The $B$-meson signal yields from the fits to the individual
bins are plotted $vs$ $M(\omega\jp)$
in Figs.~\ref{fig:slice_fits}(a) and (b).
An enhancement is evident around 
$M(\omega\jp)=3940$~MeV.
The curve in  Fig.~\ref{fig:slice_fits}(a)
is the result of a fit with a
threshold function of the
form $f(M) = A_0 q^*(M)$,
where $q^*(M)$ is the momentum of the daughter
particles in the $\omega \jp$ restframe.
This functional form accurately reproduces
the threshold behavior of Monte Carlo simulated  
$B\rt K\omega\jp$ events that are generated 
uniformly distributed over phase-space.
The fit quality to the observed data points
is poor ($\chi^2/d.o.f. = 115/11$),
indicating a significant deviation from phase-space;  
the integral of $f(M)$ over the first three 
bins is 16.8 events, where the data total is 55.6 events.

In Fig.~\ref{fig:slice_fits}(b) we show the results
of a fit where we include an $S$-wave Breit-Wigner
(BW) function~\cite{BW_swave} to represent 
the enhancement. The fit,
which has $\chi^2/d.o.f. = 15.6/8$ (CL = 4.8\%), 
yields a Breit-Wigner signal yield of $58\pm 11$ events with 
mass $M =    3943\pm 11$~MeV and width 
$\Gamma  =     87\pm 22$~MeV
(statistical errors only).
The statistical significance
of the signal, determined from
$\sqrt{-2\ln({\mathcal L}_0/{\mathcal L}_{\rm max})}$,
where ${\mathcal L}_{\rm max}$ and ${\mathcal L}_0$ are the likelihood
values for the best-fit and for zero-signal-yield, respectively, 
is $8.1\sigma$.

The $K\omega$ invariant mass distribution for
$\Mbc$-$\DE$ signal region events in the
region of the $M(\omega\jp)$ enhancement are
distributed uniformly across the available phase space
and there is no evident $K\omega$ mass structure
that might be producing the observed mass enhancement
by a kinematic reflection.  Nevertheless, the 
possibility that different high-mass $K\omega$ partial 
waves might interfere in a way that produces some 
peaking in the $\omega\jp$ mass distribution cannot 
be ruled out.

The $M(\pipi\pi^0)$ distributions
for different $M(\omega\jp)$ mass regions
exhibit $\omega\rt\pipi\pi^0$ signals that
track the $\Mbc$-$\DE$ signal yields.
The $\omega$ signal strength 
is used to infer that $(90\pm 18)\%$ of
the $B\rt K\pipi\pi^0\jp$ events 
in the $M=3943$~MeV enhancement
are produced via $\omega\rt\pipi\pi^0$ decays.

We study potential systematic errors
on the yield, mass and width by repeating the fits
with different signal parameterizations, background 
shapes and bin sizes.  For example,
when we change the background function to 
include terms up to third order in
$q^{*}$, the yield increases to $75\pm 10$ events, 
the  mass changes to $3948\pm 9$~MeV, the
width changes to $\Gamma = 100\pm 23$~MeV and the
fit quality improves: $\chi^2/d.o.f. = 10.0/6$ (CL=12.4\%).
However, the resulting background shape is very different
from that of phase-space. For different bin sizes, fitting ranges,
$M(K\omega)$ requirements, and signal line shapes we see similar variations.   

For the systematic uncertainties we use the largest
deviations from the nominal values 
for the different  fits. 
In the following, we assume that all of the $3\pi$ systems
are due to $\omega\rt\pipi\pi^0$ decays and 
include the possibility of a non-resonant contribution
in the systematic error. This is the main component
of the negative-side systematic error; the change in
yield for different background shapes contributes
a positive-side error of comparable size.
The effects of possible acceptance variation
as a function of $M(\omega\jp)$
on the mass and width values are found to be 
negligibly small.


To determine a branching fraction, we use the
BW fit shown in  Fig.~\ref{fig:slice_fits}(b)
to establish the event yield of the observed
enhancement.  
Monte Carlo simulation is used to estimate detection
efficiencies of $2.4\pm 0.1$\% and $0.42\pm 0.02$\% for 
$B\rt K^+\omega\jp$ and $K^0\omega\jp$, respectively.  
We find a product branching fraction (here we
denote the enhancement as $Y(3940)$)
\begin{equation}
{\cal B}(B\rt K Y(3940)){\cal B}(Y(3940)\rt\omega\jp) = 
(7.1 \pm 1.3 \pm 3.1)\times 10^{-5},
\end{equation}
where the second error is systematic.  The latter
includes uncertainties in the acceptance,
and the shape of the function used to
parameterize the background
and the possibility of a non-$\omega$
component of the $\pipi\pi^0$ system added in quadrature.
Here we have assumed that charged and neutral
$B$ mesons are produced in equal
numbers at the $\Upsilon(4S)$ 
and they have the same branching fractions to the 
observed enhancement~\cite{chrg-neut}.


In summary, we have observed a strong near-threshold
enhancement in the $\omega\jp$ mass spectrum
in exclusive $B\rt K\omega\jp$ decays. 
The enhancement peaks well above threshold
and is broad~\cite{resol}: if treated as an $S$-wave BW resonance, 
we find a mass of $3943\pm 11{\rm (stat)}\pm 13{\rm (syst)}$~MeV 
and a total width  $\Gamma = 87 \pm 22{\rm (stat)} \pm 26 {\rm (syst)}$~MeV.  
It is expected that a $c\bar{c}$
charmonium meson with this mass would dominantly
decay to  $D\bar{D}$ and/or $D\bar{D}^*$;  hadronic
charmonium transitions should have minuscule branching
fractions~\cite{quigg}.    

The peak mass of the observed enhancement is very similar to that 
of a peak observed by
Belle in the $\jp$ recoil mass spectrum for inclusive
$e^+e^-\rt \jp X$ events near $\sqrt{s}=10.56$~GeV~\cite{pakhlov}.
This latter peak is also seen to decay to $D\bar{D}^*$, and a search for 
it
in the $\omega\jp$ channel is in progress.  In addition, we
are examining $B\rt K D\bar{D}^*$
decays for a $D\bar{D}^*$ component of 
the enhancement reported here.

The properties of the observed enhancement are similar to 
those of some of the $c\bar{c}$-$gluon$  hybrid charmonium 
states that were first predicted in 1978~\cite{mandula} and 
are expected to be produced in $B$ meson decays~\cite{falk}.  
It has been shown that a general property of these hybrid 
states is that their decays to $D^{(*)}\bar{D}^{(*)}$ meson 
pairs are forbidden or suppressed, and the relevant ``open 
charm'' threshold is  
$m_D +m_{D^{**}} \simeq 4285$~MeV~\cite{isgur,page},  
where $D^{**}$ refers to the $J^P = (0,1,2)^+$ charmed mesons.
Thus, a hybrid state with a mass equal to that of the peak 
we observe would have large branching fractions for decays
to $\jp$ or $\psip$ plus light hadrons~\cite{close}.  
Moreover, lattice QCD calculations have indicated 
that partial widths for such decays can be comparable 
to the width that we measure~\cite{michael}.  However,
these calculations predict masses for these states
that are between 4300 and 4500~MeV~\cite{banner}, 
substantially higher than our measured value.

\section*{Acknowledgments}
We thank the KEKB group for the excellent operation of the
accelerator, the KEK cryogenics group for the efficient
operation of the solenoid, and the KEK computer group and
the NII for valuable computing and Super-SINET network
support.  We acknowledge support from MEXT and JSPS (Japan);
ARC and DEST (Australia); NSFC (contract No.~10175071,
China); DST (India); the BK21 program of MOEHRD and the CHEP
SRC program of KOSEF (Korea); KBN (contract No.~2P03B 01324,
Poland); MIST (Russia); MESS (Slovenia); Swiss NSF; NSC and MOE
(Taiwan); and DOE (USA).


\begin{thebibliography}{99}
\bibitem{conj} The inclusion of charge-conjugate modes
is always implied.
\bibitem{BW_swave} We use the form 
$\frac{\Gamma (q^*/q_0)}{(M^2-M_0^2)^2 + (M_0\Gamma (q^*/q_0))^2}$,
where $M_0$ is the peak mass and $q_0 = q^*(M=M_0)$.
\bibitem{chrg-neut}  The signal yields for the separate $K^+\omega\jp$ and
$K_S\omega\jp$ channels are consistent.
\bibitem{resol} The $M(\omega\jp)$ mass resolution is $\simeq$6~MeV
and much narrower than the observed enhancement.
\end{thebibliography}
\end{document}